\documentclass[twocolumn]{aa}  
\usepackage{epsfig}
\def\simlt{\ \raise -2.truept\hbox{\rlap{\hbox{$\sim$}}\raise5.truept   %
\hbox{$<$}\ }}
\def\simgt{\ \raise -2.truept\hbox{\rlap{\hbox{$\sim$}}\raise5.truept   %
\hbox{$>$}\ }}                                                          %

\def\be{\begin{equation}}
\def\ee{\end{equation}}
\def\newline{\hfil\break}

\def\la{\mathrel{\hbox{\rlap{\hbox{\lower4pt\hbox{$\sim$}}}\hbox{$<$}}}}
\def\ga{\mathrel{\hbox{\rlap{\hbox{\lower4pt\hbox{$\sim$}}}\hbox{$>$}}}}

\def\kpc{{\rm\,kpc}}

\def\msun{{\rm\,M_\odot}}

\begin{document}
\title{On the DM interpretation of the origin of non-thermal phenomena in galaxy clusters}
   \author{S. Colafrancesco\inst{1}, R. Lieu\inst{2}, P. Marchegiani\inst{1,3}, M. Pato\inst{4,5,6}
   and L. Pieri\inst{7,8}}
   \offprints{S. Colafrancesco}
   \institute{   ASI-ASDC
              c/o ESRIN, Via G. Galilei snc, I-00040 Frascati, Italy
              Email: Sergio.Colafrancesco@asi.it
   \and
              The University of Alabama Huntsville, Optics Building, Room 201 F
              Email:richardlieuuah@gmail.com
   \and
              Dipartimento di Fisica, Universit\`a di Roma La Sapienza, P.le A. Moro 2, Roma, Italy
              Email: marchegiani@mporzio.astro.it
   \and
             Dipartimento di Fisica, Universit\`a degli Studi di Padova, via Marzolo 8, I-35131, Padova, Italy
   \and
             Institut d'Astrophysique de Paris, 98bis bd Arago, 75014, Paris, France
   \and
             Universit\'e Paris Diderot-Paris 7, rue Alice
             Domon et L\'eonie Duquet 10, 75205, Paris, France
             Email:pato@iap.fr
   \and
             Istituto Nazionale di Fisica Nucleare, Sezione di Padova - via Marzolo 8, I-35131, Padova, Italy
             Email:lidia.pieri@gmail.com
   \and
             Dipartimento di Astronomia, Universit\`a degli Studi di Padova, Vicolo dell'Osservatorio 3, I-35122 Padova
             }
\date{Received  / Accepted  }
\authorrunning {S. Colafrancesco et al.}
\titlerunning {On the DM interpretation of non-thermal phenomena in galaxy clusters}
%
\abstract
  {}
   {
   We studied the multi-frequency predictions of various annihilating Dark Matter (DM) scenarios in order
   to explore the possibility to interpret the still unveiled origin of non-thermal phenomena
   in galaxy clusters.}
  {
  We consider three different DM models with light (9 GeV), intermediate (60 GeV)
  and high (500 GeV) neutralino mass and we study their physical effects in the atmosphere of the
  Coma cluster. The secondary particles created in the neutralino annihilation processes
  produce a multi-frequency Spectral Energy Distribution (SED) of non-thermal radiation, as well
  as heating of the intracluster gas, that are tested against the
  observations available for the Coma cluster, from radio to
  gamma-rays. The various DM produced SEDs are normalized by the condition to fit the Coma
  radio halo spectrum, thus obtaining best-fit values of the
  annihilation cross-section $\sigma V$ and of the central magnetic
  field $B_0$.
  }
   {
  We find that it is not possible to interpret all the non-thermal
  phenomena observed in galaxy clusters in terms of DM annihilation.
  The light mass DM model with 9 GeV mass produces too small power at all
  other frequencies, while the high mass DM model with
  500 GeV produces a large excess power at all other frequencies.
  The intermediate mass DM model with 60 GeV and $\tau^{\pm}$ composition
  is marginally consistent with the HXR and gamma-ray observations but
  barely fails to reproduce the EUV and soft X-ray observations.
  The intermediate mass DM model with 60 GeV and $b{\bar b}$
  composition is, on the other hand, always below the observed fluxes.
  We note that the radio halo spectrum of Coma is well fitted only in the $b{\bar b}$
  or light and intermediate mass DM models.
  We also find that the heating produced by the DM annihilation in the
  center of the Coma cluster is always larger than the
  intracluster gas cooling rate for an NFW DM density profile and
  it is substantially smaller than the cooling rate only for a cored DM
  density profile in light mass DM model with 9 GeV.
  }
   {
We conclude that the possibility of interpreting the origin of
non-thermal phenomena in galaxy clusters with DM annihilation
scenarios requires a low neutralino mass and a cored DM density
profile. If we then consider the multimessenger constraints to the
neutralino annihilation cross-section, it turns out that such
scenario would also be excluded unless we introduce a substantial
boost factor due to the presence of DM substructures. }


 \maketitle

\section{Introduction}
 \label{sec.intro}

Observational evidence for the existence of diffuse
components of non-thermal origin in clusters of galaxies and their wider implications is
a topic of sustained debate among researchers.
In particular, the presence of giant radio halos and relics (see
Feretti \& Giovannini 2008 for a review) indicates that diffuse
distributions of relativistic electrons and magnetic fields must
exist in many clusters of galaxies.
Moreover, some clusters show also the presence of `unusual'
diffuse components, i.e. the excess soft and hard X-ray emission
above the level expected from the hot virialized cluster (e.g.,
Lieu et al 1996, Fusco-Femiano et al 1999, Kaastra et al 1999),
even if sometimes the observations are controversial (e.g.
Rossetti \& Molendi 2004).

While the diffuse radio sources in galaxy clusters are
universally interpreted as produced by synchrotron emission,
the Extreme Ultraviolet (EUV) and
Hard X-ray (HXR) excesses have been interpreted
as the combined effect of an underlying
inverse-Compton scattering (ICS) of the Cosmic Microwave Background (CMB)
against cluster relativistic electrons that complement the radio
halo population (e.g., Hwang 1997, Ensslin \& Biermann 1998, Sarazin \& Lieu
1998, Blasi \& Colafrancesco 1999, Colafrancesco et al. 2005,
Marchegiani, Perola \& Colafrancesco 2007, Colafrancesco \&
Marchegiani 2009, Buote 2001), or thermal emission (Mittaz et
al 1998, Cen \& Ostriker 1999) from `missing baryons' at
sub-virial temperatures (see, e.g., the review of Durret et al
2008).
Focusing upon the former interpretation, the role and
ramifications of an underlying power law spectrum across the
entire X-ray frequency band can be far reaching, with issues
concerning cluster merging and central cooling, and also mass and
spectroscopic consequences being addressed by Million \& Allen
(2009), Lagana et al (2010), and Prokhorov (2009).
In the even broader context, the observed cosmic $^6$Li abundance
could be evidence for excess entropy in clusters (Nath et al.
2006), along with the surprisingly low level of Sunyaev-Zel'dovich
effect in the WMAP and SPT sample of many clusters (Diego and
Partridge 2010, Lieu et al. 2006, Komatsu et al. 2010, Lueker et
al. 2009), and the effects of the magnetic field on the
intracluster gas distribution (e.g. Colafrancesco \& Giordano
2007) as possible indication of significant intracluster
non-thermal pressure.

Historically however, there have been numerous suggestions of a
general, non-thermal intracluster environment, early examples
being Jaffe (1977) and Rephaeli (1979) who considered Alfven wave
excitation. In particular, a model of the acceleration via Alfven
waves driven by major cluster mergers is given in Brunetti et al.
(2004), including the idea that relativistic jets carry a
significant portion of the total energy output of radio galaxies,
causing X-ray emission up to Mpc distance scales (Ghisellini \&
Celotti 2001). Celotti et al. (2001) provided an analysis of such
a scenario for the case of PKS 0637-752.\\
A relevant, related observation (see Giovannini et al. 1993) maps
the spectral index of cluster radio halos, and specifically of the
Coma radio halo from 327 MHz to 1.4 GHz. Here, we note the need
for non-thermal emission by electrons with energy just above the
energy range required by the non-thermal electron population
supposed to produce soft and hard X-rays by ICS (see Lieu et al.
1999).

An important issue connected with the presence of diffuse emission
of non-thermal origin in galaxy clusters is that relativistic
electrons with $E\sim$ GeV (as those which produce the radio halos)
lose their energy in relatively short time by effect of synchrotron
and ICS with the CMB photons interactions,
and they should reach a short distance in comparison to the dimension
of the halos (see discussion in Blasi \& Colafrancesco 1999).
Therefore, several solutions have been proposed to solve this
problem, as the re-acceleration \textit{in situ} of the electrons
by Alfven wave excitation driven by major cluster mergers and
induced turbulences (e.g., Jaffe 1977, Rephaeli 1979, Brunetti et
al. 2004) or the production of secondary electrons in the
interactions between non-thermal and thermal protons (e.g.,
Dennison 1980, Blasi \& Colafrancesco 1999, Marchegiani et al.
2007, Colafrancesco \& Marchegiani 2008).\\

The separate physical mechanism to be invoked is neutralino Dark
Matter (DM) annihilation as a cluster's non-thermal reservoir (see
Colafrancesco \& Mele 2001, Colafranceso, Profumo, \& Ullio 2006)
to compensate for losses of the particles as they are distributed
throughout the cluster medium during the long process of particle
diffusion (see, e.g., V\"olk 1996, Colafrancesco \& Blasi 1998).

Today, several astrophysical evidences (e.g., gravitational
lensing, galaxy rotation curves, galaxy clusters masses) indicate
that most of the matter content of the universe is in form of DM,
whose nature is still elusive. Several candidates have been
proposed as fundamental DM constituents, ranging from axions to
light, MeV DM, from Kaluza-Klein (KK) particles, branons,
primordial Black Holes (PBH), mirror matter to supersymmetric
Weakly Interacting Massive Particles (WIMPs; see, e.g., Baltz
2004, Bertone et al. 2005, and Bergstr\"om 2000 for reviews). In
this paper we will assume that the main DM constituent is the
lightest neutralino of the minimal supersymmetric extension of the
Standard Model (MSSM). This assumption can be tested by observing
the radiation emitted by the secondary products of the neutralino
annihilation processes, whose features have been studied (e.g.,
Colafrancesco \& Mele 2001, Colafrancesco et al. 2006). As pointed
by Colafrancesco et al. (2006), the relevant physical properties
which determine the features of the emitted radiation are the
composition of the neutralino, its mass, and the value of the
annihilation cross section.

In this paper we explore the consequences of a neutralino DM
annihilation scenario on the non-thermal particle content of
galaxy clusters over a multi-frequency range (from radio to
gamma-ray and TeV energies), and in particular on one of the best
studied clusters so far, like the Coma cluster of galaxies, for
which an extensive multi-frequency coverage has been
accumulated.\\
To this aim, we focus on three neutralino DM models with mass
$M_\chi$ of 9, 60 and 500 GeV.

Neutralinos with 60 and 500 GeV mass are representative of,
respectively, intermediate and high mass supersymmetric models
which are consistent with accelerator and multi-messenger
constraints, and that are also interesting to be probed by direct
detection experiment (like e.g. the recent CDMS results, see Ahmed
et al. 2009).
In this context, the 60 GeV neutralino mass is close to the lower
limit of the combination of the CDMS II experiment with the
allowed parameter space calculated from various Minimal
Supersymmetric Models (e.g., Roszkowski et al. 2007, Ellis et al.
2005), and it is interesting for being possibly tested by a
multi-frequency search limited to the gamma-ray energies probed by
Fermi.
On the other side, the 500 GeV mass is interesting for being
tested by a multi-frequency search extended even in Cherenkov
experiments realm.

The light mass neutralino model with $M_{\chi}=9$ GeV has been
recently worked out as a consistent solution of the CoGeNT and
DAMA observations (Fitzpatrick et al. 2010). Such a low mass
neutralino is allowed in supersymmetric scenarios where the
hypothesis of gaugino masses universality at the unification scale
is released (Bottino et al. 2003).

To explore the astrophysical consequences of such DM models, we
consider pure annihilation channels in order to highlight their
observational features.
We focus, specifically, on $b {\bar b}$ and $\tau^{\pm}$
annihilations for 9, 60 and 500 GeV mass neutralino models, and we
also consider annihilations into $W^{\pm}$ for the 500 GeV mass
neutralino model. We leave the neutralino annihilation
cross-section $\sigma V$ as a free parameter to be constrained, in
our analysis, by the astrophysical data.\\
In this study, we also consider the constraints on the DM model
and density profile coming from the heating rate of the thermal
gas at the centre of the cluster.

The outline of the paper is the following. We describe in Sect.2
the properties of the various neutralino DM models and the various
DM density profiles that we consider for the Coma cluster.
We derive the astrophysical consequences of the neutralino DM
annihilation in Coma in Sect. 3, and we focus on the limits set on
the expected DM-induced spectral energy distribution (SED) at
multifrequency, as well as on the additional physical effects
induced by DM annihilation. The discussion of our results and the
conclusion of our analysis are presented in Sect.4.

Throughout the paper, we use a flat, vacuum--dominated
cosmological model with $\Omega_m = 0.3$, $\Omega_{\Lambda} = 0.7$
and $H_0 = 70$ km s$^{-1}$ Mpc$^{-1}$.

\section{The Dark Matter model}

A 9 or 60 GeV supersymmetric particle is expected to annihilate in
quarks and leptons. Direct annihilation into light fermions is
helicity suppressed, so that the main remaining channels of
annihilations are $b \bar b$ and $\tau^+ \tau^-$. Electrons and
positrons will be, then, mainly produced through secondary
processes of hadronization and/or decay. Direct annihilation into
$\gamma$s happens at 1-loop level and is therefore highly
suppressed too. The bulk of photons arises from the $\pi^0 \to
\gamma \gamma$ e.m. decay.\\
In addition to the aforementioned channels, a 500 GeV supersymmetric
particle can also annihilate into gauge bosons, heavy quarks and Higgs
particles. The yield of photons produced in these cases, as well as their
energy spectrum, is very much similar to each other, so that we choose
as a representative case the annihilation into $W^+W^-$.

In this framework, we have computed the positron/electron spectra
with DarkSUSY (Gondolo et al. 2004) and the yield of gamma-ray
photons according to Fornengo, Pieri \& Scopel (2004).

\subsection{The Dark Matter profile of the Coma cluster}

Recent determinations of the mass and density profile of Coma
include methods based on weak lensing (Kubo et al. 2007) as well
as on the study of the velocity moments of early-type galaxies in
the Coma cluster (Lokas \& Mamon 2003) or on the analysis of
infall patterns around Coma (Geller et al. 1999).
Numerical simulations also give a hint on the DM density profile
inside clusters (Bullock et al. 2001). The results of astronomical
measurements and numerical experiments lead to a determination of
the mass profile which is consistent among the various methods,
within the errors and the spatial resolution of each approach. In
this paper we use the results of Lokas \& Mamon (2003), since they
provide a fit to the data with both a spiky NFW profile and a
cored profile.

The total inferred dark matter mass of Coma is $M_{vir}=1.2 \cdot 10^{15} \msun$
within a dark matter virial radius of 2.7 Mpc.
The profile is parametrized as:
\begin{equation}
\rho(r) =\frac{\rho_s}{\left ( \frac{r}{r_s} \right )^\alpha
\left(1+\frac{r}{r_s}\right)^{3-\alpha}} . \label{rho}
\end{equation}
Scale parameters and inner slope can be found in Table
\ref{tab:par}. Also shown are the parameters for an NFW profile as
inferred by N-body simulations (Bullock et al. 2001).
\begin{table}
\begin{center}
\begin{tabular}{cccc}
profile & $\alpha$ & $r_s (\kpc)$ & $\rho_s \over \rho_c$  \\
\hline NFW & 1 & 287 & $1.96 \times 10^{4}$ \\ cored & 0 & 142 &
$1.66 \times 10^{5}$ \\ NFW$^{N-body}$ & 1 & 450 & $6.75 \times
10^{3}$ \\ \hline
\end{tabular}
\caption{Density profiles for the Coma cluster. $\rho_c = 1.36\times10^2
\msun \kpc^{-3}$ is the critical density for closure of the
universe.} \label{tab:par}
\end{center}
\end{table}
\begin{figure}[ht]
\begin{center}
\epsfig{file=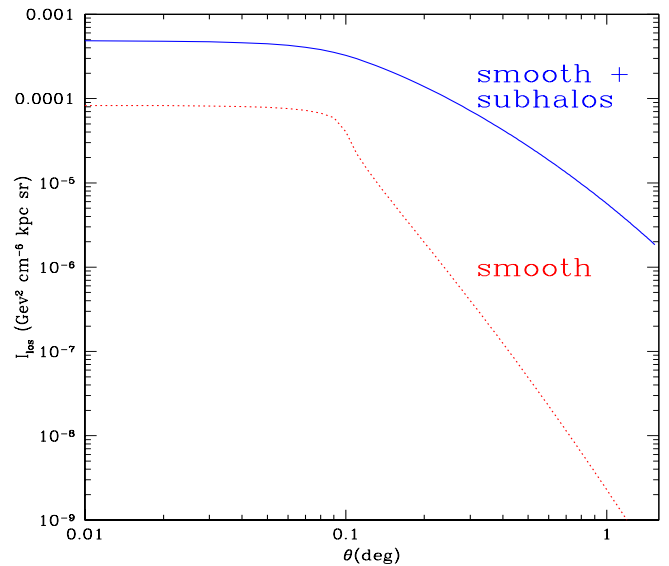,height=8.cm,width=9.cm,angle=0.0}
\end{center}
 \caption{\footnotesize{
The los integral of the DM density squared is shown as a function
of the projected angle from the cluster center for a resolution of
0.1 degree (corresponding to a solid angle of $10^{-5}$ sr). The
red dotted curve is for a smooth DM density profile while the blue
solid curve is for a smooth DM density profile with the addition
of subhalos (see text for details).
 }}
 \label{Coma_boost}
\end{figure}

We also compute the effective boost factor seen as a line of sight
(los) integral  due to the presence of subhalos orbiting in the
gravitational potential of the Coma cluster. We follow the
methodology and results of Pieri et al. (2009) in the framework
suggested by the Via Lactea II numerical simulation (Diemand et
al. 2008).\\
We make therefore the following assumptions:
i) the slope of the subhalo mass function is -2, which implies,
for a DM halo such as Coma, that about 64\% of its mass is
virialized in subhalos;
ii) the mass of subhalos is in the range $10^{-6} M_\odot -
10^{-2} M_{vir}$;
iii) the radial dependence of the subhalo concentration parameter;
iv) an NFW profile for each of the subhalos.\\
We then compute the los integral, with and without subhalos, of
the DM density squared in order to infer the boost factor ${\cal
B}$ defined as the integral along the line of sight and on the
solid angle of the neutralino density squared in the case in which
we consider a smooth DM density profile or the sum of smooth DM
density plus DM subhalos (see, e.g., Pieri et al. 2009).
Fitting formulas for the concentration parameter as well as
explicit notation for the line of sight integral including a
population of subhalos can be found in Pieri et al. (2009). \\
Fig.\ref{Coma_boost} shows the result of our computation in the
case of an angular resolution $\Delta \Omega = 10^{-5}$ sr
(corresponding to 0.1 degree). Once integrating over the whole
cluster volume, we obtain a boost factor  ${\cal B} = 35$. We do
not expect significant deviation from this value when changing the
subhalo profile.

In the following calculations we will first consider the case of a
smooth DM density profile and then the effects of the boost
factor.

\section{Astrophysical consequences}

We discuss here various astrophysical consequences of the DM models
we consider in this paper: e.m. signals (all of non-thermal
origin) and heating of the intracluster gas induced by secondary
particles produced in the DM annihilation processes.

The complete description of the emission features induced by DM
annihilation is done by using the diffusion equation (i.e.
neglecting convection and re-acceleration effects) given in
Colafrancesco, Profumo \& Ullio (2006, see their eq. 35) which
takes into account, consistently, the diffusion and energy loss
properties of the secondary particles produced in the neutralino
DM annihilation process.

The energy loss term is the sum of effects due to Inverse Compton,
synchrotron radiation, Coulomb losses and Bremsstrahlung (see eq.
A.16 in Colafrancesco et al. 2006 for details). For electron
energies $\simgt$ GeV the Inverse Compton and synchrotron terms
dominate the losses while for energies below $\sim 140$ MeV,
Coulomb losses dominate (see Fig. A.3 in Colafrancesco et al.
2006).
For a more detailed description of the energy losses in the Coma
cluster, we use throughout the paper a B-field radial profile
$B(r) \propto n_{th}(r)$, where $n_{th}$ is the intracluster gas
number density, unless otherwise specified.

The derivation of the full solution of the diffusion equation and
the effects of diffusion and energy losses are given in
Colafrancesco et al. (2006, see Eq. A.1) and we refer to this
paper for technical details. We consider here the same analytical
approach and we present the results of the multi-frequency
emission produced by the DM models under study
for the Coma cluster.
We also compute the DM-induced
signals in the limit in which electrons and positrons lose energy
on a timescale much shorter than the timescale for spatial
diffusion, i.e. the regime which applies to the case of galaxy
clusters (see Colafrancesco et al. 2006 for a discussion).

\begin{figure}[ht]
\begin{center}
{
 \epsfig{file=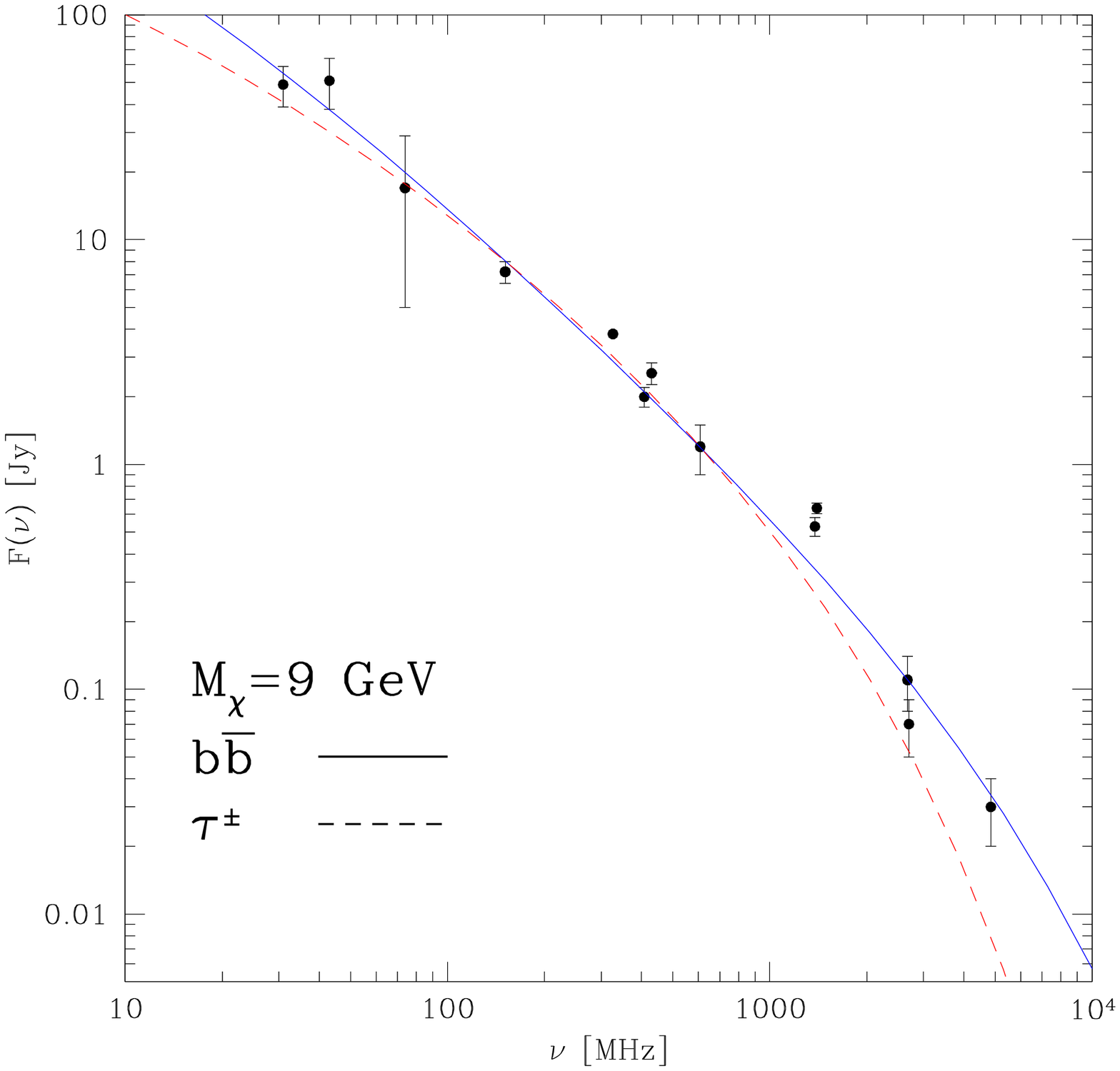,height=6.cm,width=9.cm,angle=0.0}
 \epsfig{file=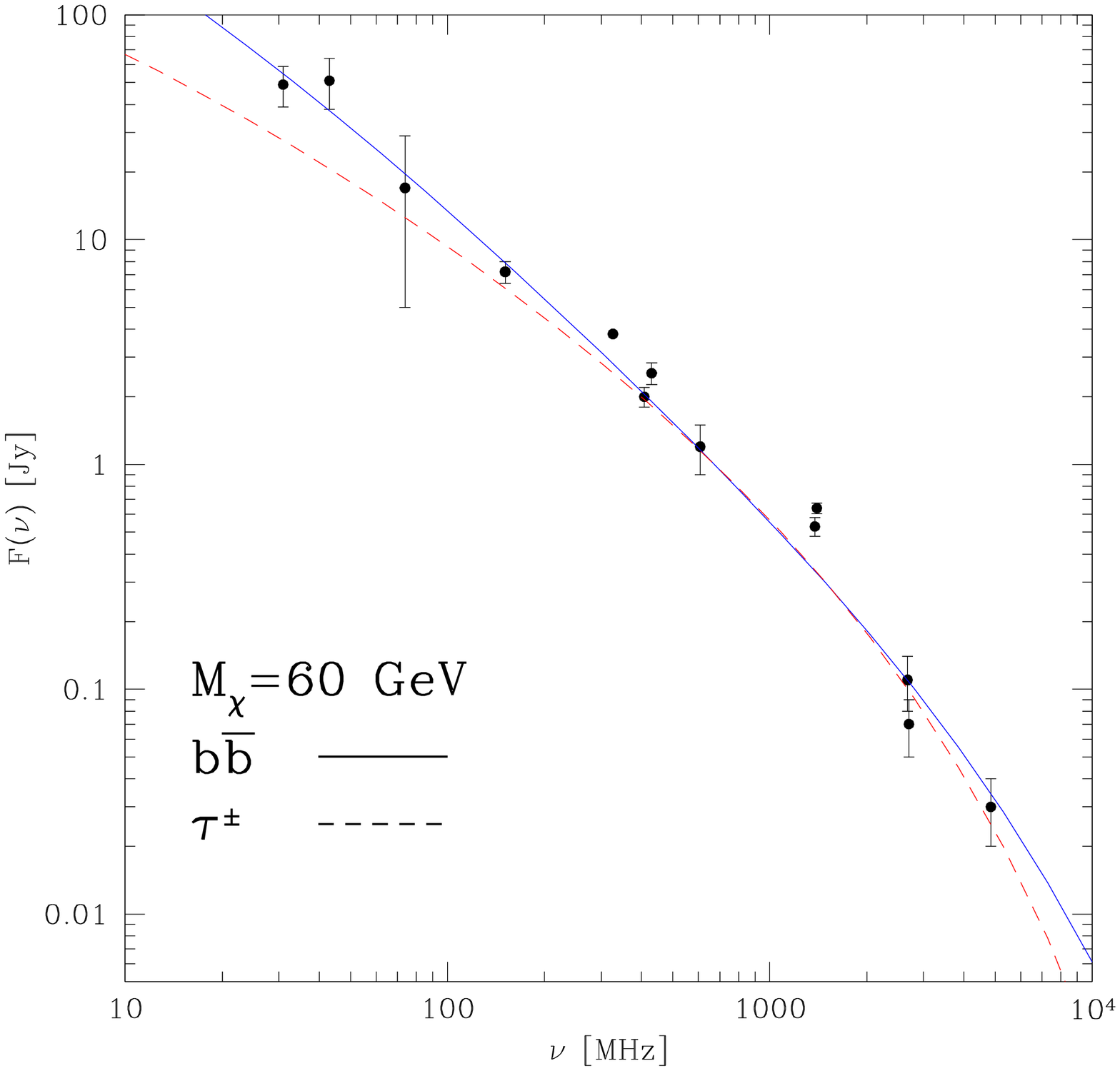,height=6.cm,width=9.cm,angle=0.0}
 \epsfig{file=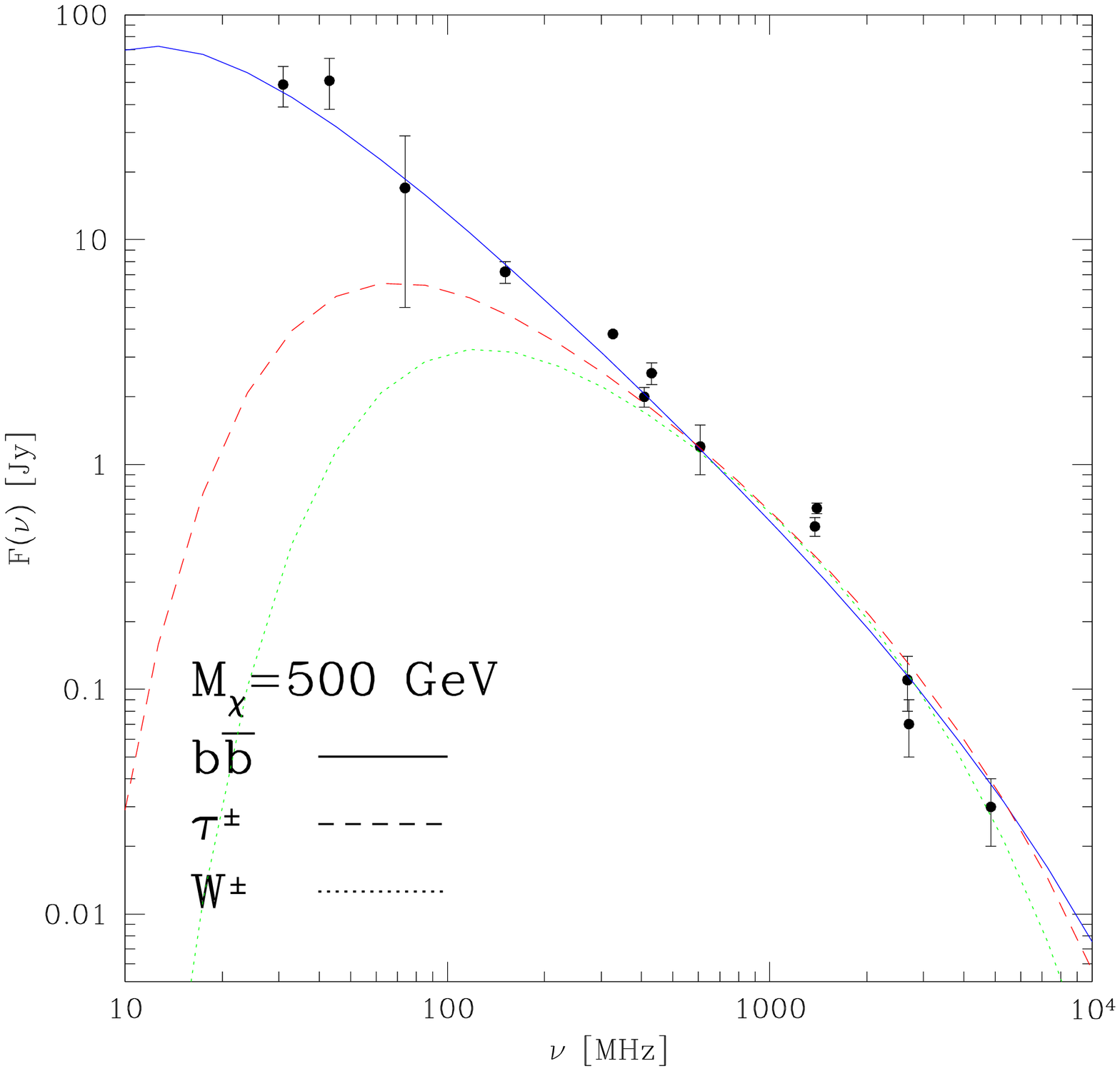,height=6.cm,width=9.cm,angle=0.0}
}
\end{center}
 \caption{\footnotesize{The radio halo spectrum of Coma and the best fits of three DM
 models with $M_{\chi}=9$ GeV (upper panel) for $b {\bar b}$ (solid with
 $\sigma V=1.7 \cdot 10^{-25}$ cm$^3$ s$^{-1}$ and $B_0= 20$ $\mu$G)
 and $\tau^{\pm}$ (dashed with $\sigma V=5.6 \cdot 10^{-25}$ cm$^3$ s$^{-1}$ and $B_0= 2.2$ $\mu$G),
 $M_{\chi}=60$ GeV (mid panel) for $b {\bar b}$ (solid with
 $\sigma V=8.5 \cdot 10^{-23}$ cm$^3$ s$^{-1}$ and $B_0= 0.5$ $\mu$G)
 and $\tau^{\pm}$ (dashed with $\sigma V=9.0 \cdot 10^{-22}$ cm$^3$ s$^{-1}$ and $B_0= 0.1$ $\mu$G),
 and with $M_{\chi}=500$ GeV (lower panel) with $b {\bar b}$
 (solid with $\sigma V= 2.0 \cdot 10^{-18}$ cm$^3$ s$^{-1}$ and $B_0= 0.01$ $\mu$G),
 $\tau^{\pm}$ (dashed with $\sigma V= 1.9 \cdot 10^{-17}$ cm$^3$ s$^{-1}$ and $B_0= 0.002$ $\mu$G)
 and W$^{\pm}$ (dotted with $\sigma V= 1.3 \cdot 10^{-16}$ cm$^3$ s$^{-1}$ and $B_0= 0.001$ $\mu$G).
 Data from Thierbach et al. (2003).
 The results are obtained using $B(r)\propto n_{th}(r)$
 and NFW smooth DM profiles, without considering the effect of substructures.
 }}
 \label{Coma_rh_bestfit}
\end{figure}

The part of the SED from which the strongest spectral constraints
can be set is the low-frequency radio band from $\sim 30$ MHz to
$\sim 5$ GHz.\\
Fig.\ref{Coma_rh_bestfit} shows the spectra of the diffuse radio
emission of Coma calculated in the various DM models that we
consider here. The best fit values of the central value of the
magnetic field $B_0$ and $\sigma V$ have been obtained from the
fit to the Coma radio halo spectrum and normalizing the curves at
the data point at 608 MHz.\\
We notice that the radio-halo spectrum could discriminate between
the different neutralino compositions, preferring a $b {\bar b}$
composition due to its better-fitting intrinsic spectral curvature
(see Fig.\ref{Coma_rh_bestfit}).

The fit is influenced by the value of the B-field. Changing the
intensity of the B-field implies to change the energies of the
electrons that emit by synchrotron in this frequency range, and
this implies, in turn, that the resulting radio halo spectrum has
a change in the spectral shape according to the value of $B_0$.
These effects are illustrated in Fig.\ref{Coma_rh_bestfit}.
For $M_\chi=9$ GeV, the $b\bar b$ model recovers quite well the
curvature of the observed spectrum for $B_0=20$ $\mu$G, while the
$\tau^\pm$ model reproduces the observed spectrum for a value of
$B_0 =2.2$ $\mu$G.
For $M_\chi=60$ GeV, the $b\bar b$ model recovers quite well the
curvature of the observed spectrum for $B_0=0.5$ $\mu$G, while the
$\tau^\pm$ model reproduces the observed spectrum only for the
frequency range $\nu>100$ MHz for a value of $B_0 =0.1$ $\mu$G.
For $M_\chi=500$ GeV, the $b\bar b$ model reproduces the curvature
of the observed spectrum for $B_0=0.01$ $\mu$G, while the
$\tau^\pm$ and $W^\pm$ models require values of the central
magnetic field that are much lower ($B_0=0.002 - 0.001$ $\mu$G)
and can reproduce the spectral shape of the radio halo only for
$\nu>400$ MHz. The corresponding best fit values of the
annihilation cross section $\sigma V$ are given in the caption of
Fig.\ref{Coma_rh_bestfit}.

\begin{figure}[ht]
\begin{center}
\epsfig{file=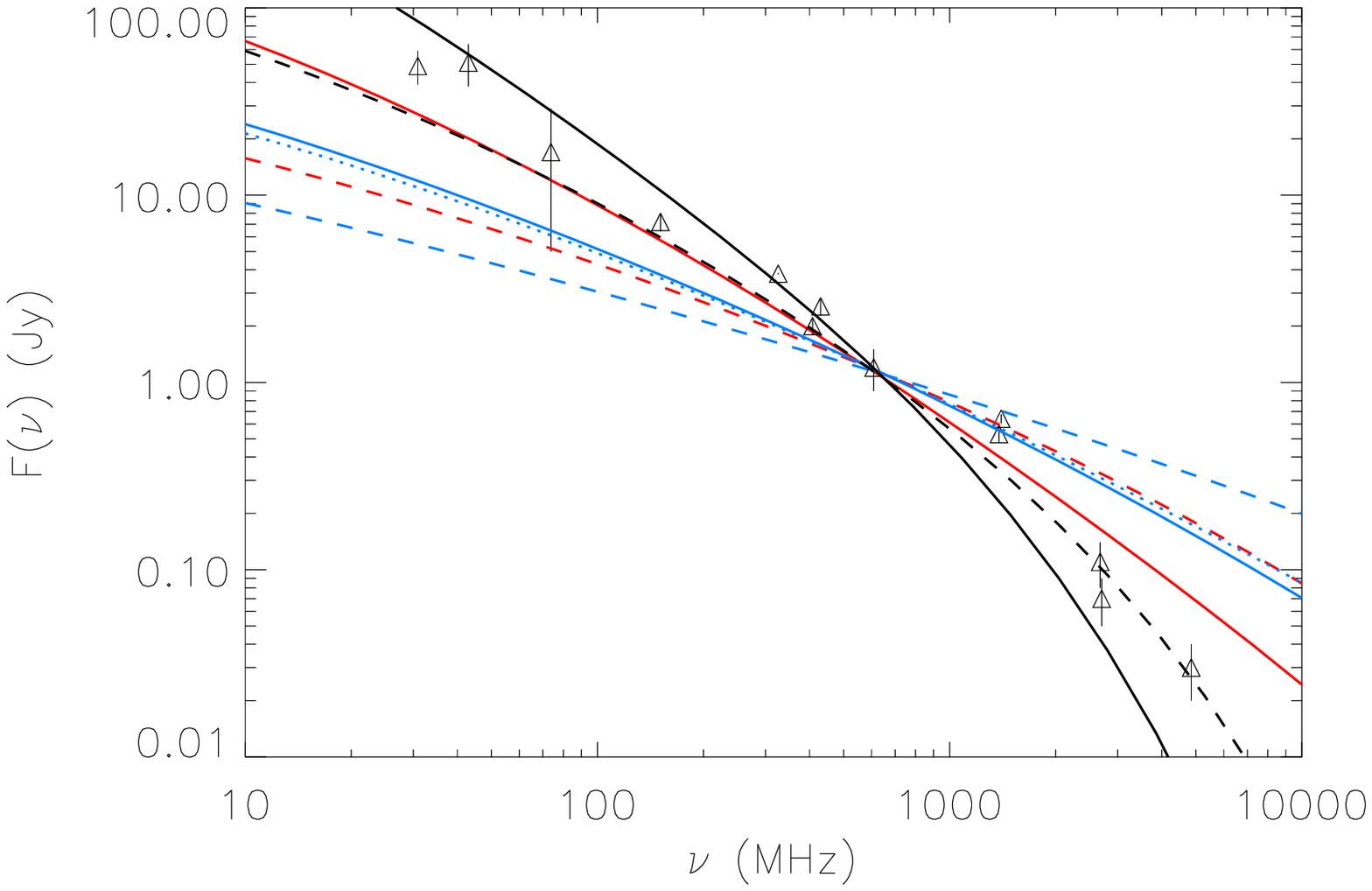,height=8.cm,width=9.cm,angle=0.0}
\end{center}
 \caption{\footnotesize{
The radio halo spectrum of Coma and the predictions of DM models
calculated by assuming $B_0=4.7$ $\mu$G and $B(r)\propto
n_{th}(r)^{0.5}$; black lines refer to $M_{\chi}=9$ GeV with
$b{\bar b}$ (solid line; $\sigma V=6.0\times10^{-25}$ cm$^3$
s$^{-1}$) and $\tau^\pm$ (dashed line; $\sigma
V=1.4\times10^{-25}$ cm$^3$ s$^{-1}$); red lines refer to
$M_{\chi}=60$ GeV with $b{\bar b}$ (solid line; $\sigma
V=1.0\times10^{-24}$ cm$^3$ s$^{-1}$) and $\tau^\pm$ (dashed line;
$\sigma V=9.0\times10^{-25}$ cm$^3$ s$^{-1}$); blue lines refer to
$M_{\chi}=500$ GeV with $b{\bar b}$ (solid line; $\sigma
V=7.5\times10^{-24}$ cm$^3$ s$^{-1}$), $\tau^\pm$ (dashed line;
$\sigma V=3.0\times10^{-23}$ cm$^3$ s$^{-1}$) and $W^\pm$ (dotted
line; $\sigma V=1.1\times10^{-23}$ cm$^3$ s$^{-1}$).
The results are obtained using NFW smooth DM profiles, without considering the effect of substructures.
 }}
 \label{Coma_rh}
\end{figure}

Fig.\ref{Coma_rh} shows the radio halo spectrum calculated in the
various DM models that we consider here where we assumed the value
of the central magnetic field derived from the most recent
analysis of Faraday Rotation measures, $B_0=4.7$ $\mu$G with a
radial profile given by $B(r)\propto n_{th}(r)^{0.5}$ (Bonafede et
al. 2010). In this figure we have fixed the value of $\sigma V$
such that the radio halo spectrum fits the observed one at  608
MHz.
For this value of the central B-field, the DM model with
$M_{\chi}=9$ GeV and $\tau^\pm$ composition and the model with
$M_{\chi}=60$ GeV and $b\bar b$ composition can (marginally)
reproduce the observed shape of the spectrum, while the model with
9 GeV and $b\bar b$ composition is too steep and all other DM
models provide radio halo spectra much flatter (harder) than the
measured one.

\begin{figure}[ht]
\begin{center}
 \epsfig{file=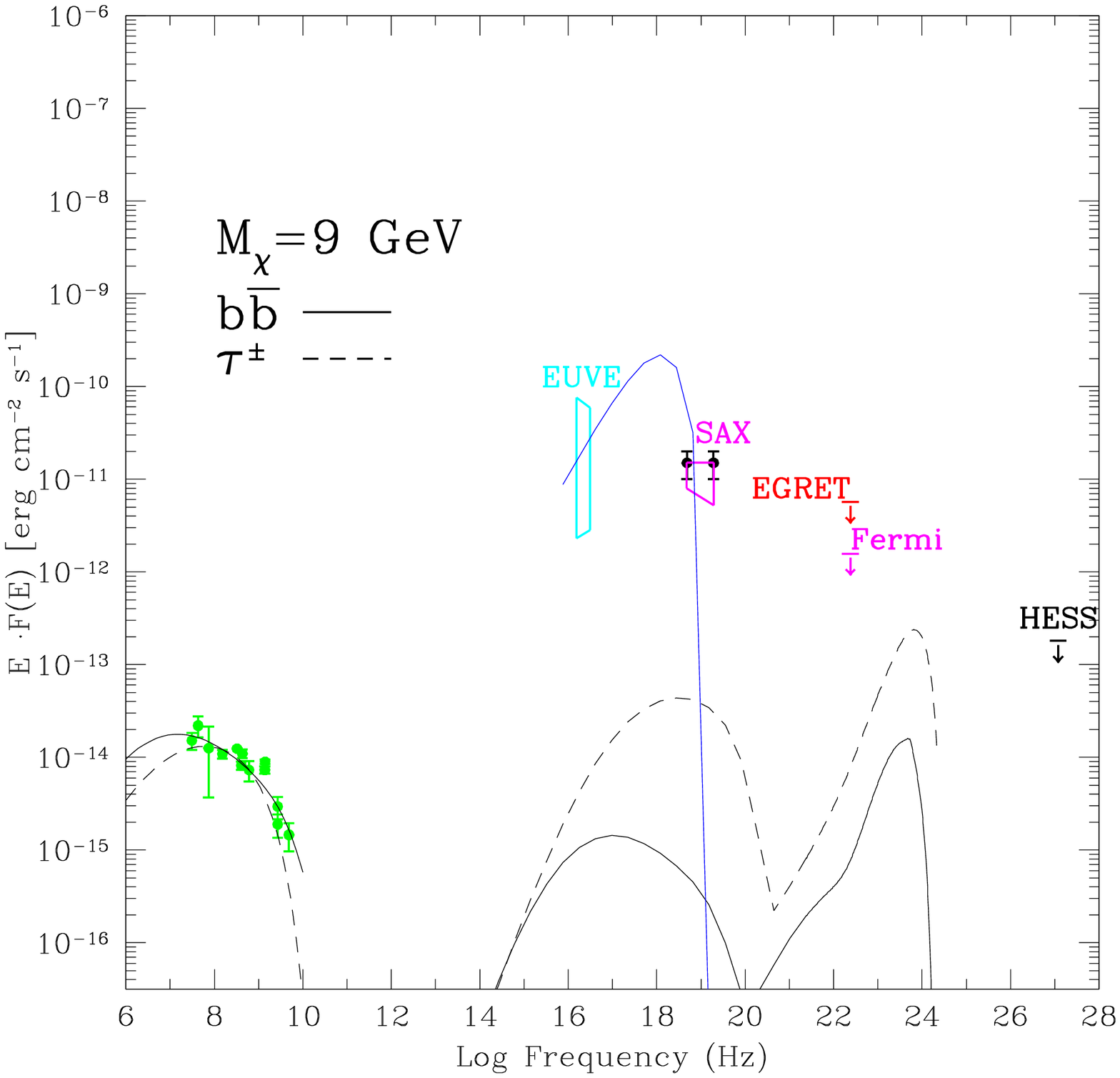,height=6.cm,width=9.cm,angle=0.0}
 \epsfig{file=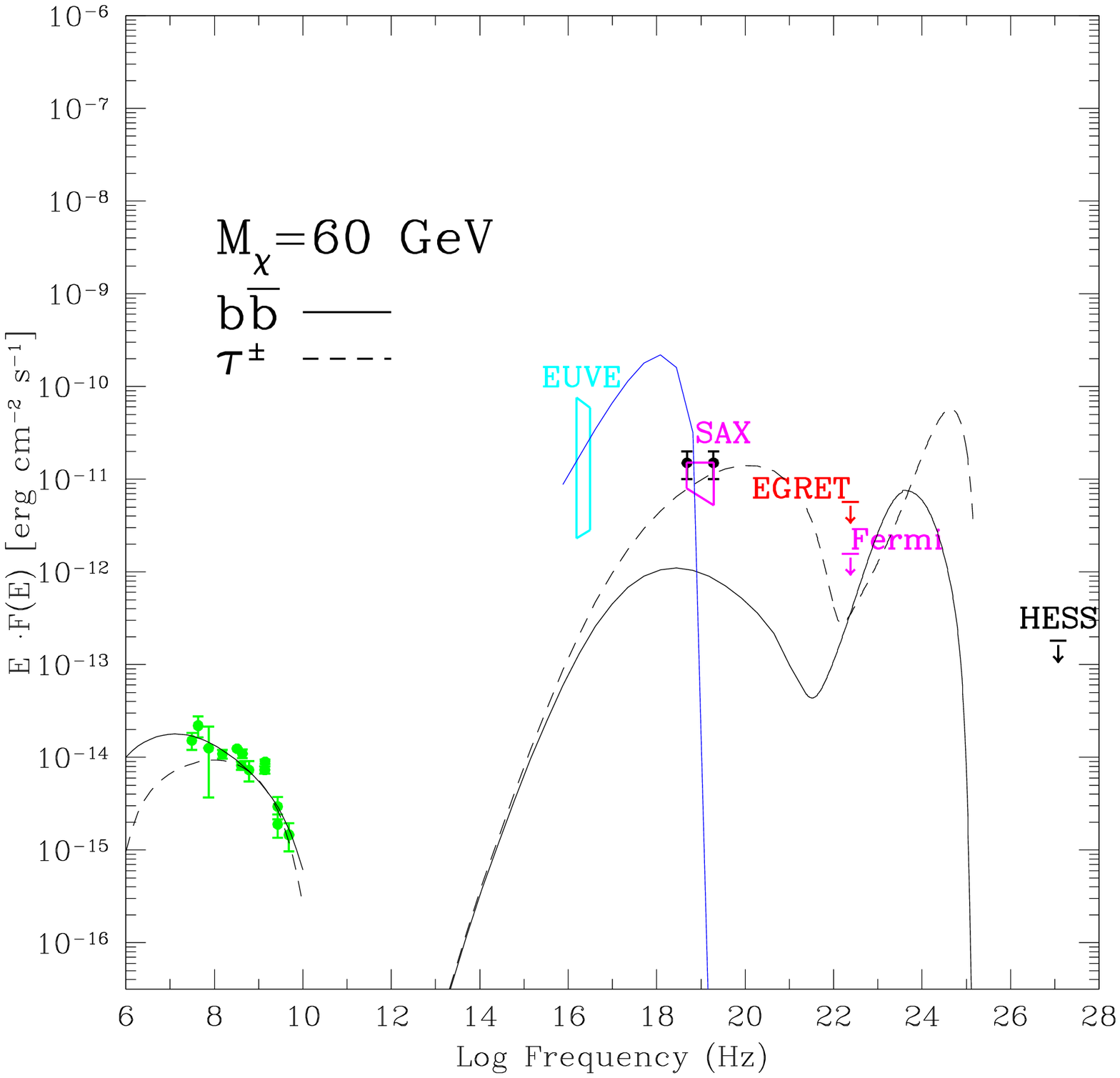,height=6.cm,width=9.cm,angle=0.0}
 \epsfig{file=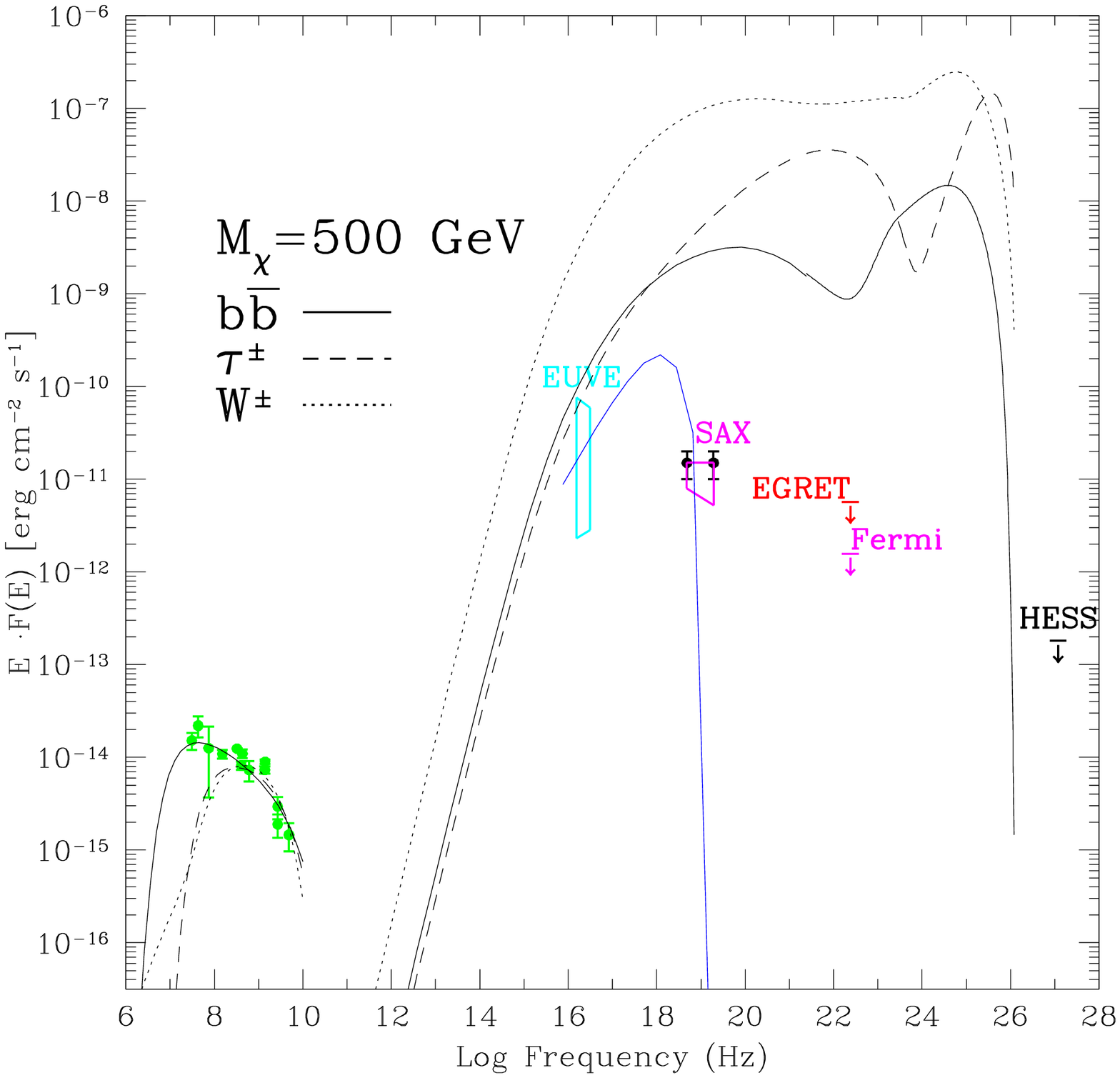,height=6.cm,width=9.cm,angle=0.0}
\end{center}
 \caption{\footnotesize{The SED of the DM-induced e.m. signals in
 Coma as predicted by the three DM models.
 Upper panel: $M_{\chi}=9$ GeV (solid: $b {\bar b}$; dashed: $\tau^{\pm}$);
 mid panel: $M_{\chi}=60$ GeV (solid: $b {\bar b}$; dashed: $\tau^{\pm}$);
 lower panel: $M_{\chi}=500$ GeV
 (solid: $b {\bar b}$; dashed: $\tau^{\pm}$; dotted: $W^{\pm}$)
 The values of $\sigma V$ and $B_0$ for the different models are
 the same used in Fig.\ref{Coma_rh_bestfit}.
 The hatched regions in the X-ray domain represent data from PDS/BeppoSAX
 (Fusco-Femiano et al. 2004), HEXTE/RXTE (Rephaeli et al. 1999) and EUVE
 (Lieu et al. 1999). We also report the EGRET (Reimer et al 2003), Fermi (Mori 2009)
 and HESS (Aharonian et al. 2009a) upper limits on Coma.
 The blue line is a thermal bremsstrahlung model for Coma with $kT=8.2$ keV
 (Briel et al. 1992).
 The results are obtained using $B(r)\propto n_{th}(r)$
 and NFW smooth DM profiles, without considering the effect of substructures.
 }}
 \label{Coma_sed}
\end{figure}

The overall spectral energy distributions (SEDs) of the DM models
derived from the fits to the Coma radio halo spectrum populate
differently the high-frequency regions of the e.m. spectrum.\\
The SEDs of a 9 GeV, of a 60 GeV and of a 500 GeV neutralino models
annihilating in Coma are shown in Fig. \ref{Coma_sed}.

The 9 and 60 GeV mass DM models that best fit the radio-halo
spectrum of Coma produce ICS and bremsstrahlung radiation and
$\pi^0 \to \gamma \gamma$ radiation consistent with the existing
limits at multi-frequency. Under the previous conditions, these
signals cannot, in particular, be the nature of the soft and hard
X-ray emission excesses detected in this cluster. Only the 60 GeV
$\tau^{\pm}$ neutralino model can fit both the radio halo spectrum
and the hard X-ray data, but not the Extreme UV Explorer (EUVE)
data.

The DM model with 500 GeV mass can reproduce the Coma radio halo
spectrum only with very low values of the central magnetic field,
$B_0 \sim 0.001 - 0.1 $ $\mu$G. This fact has the consequence of
requiring quite large values for the annihilation cross section
$\sigma V \sim 10^{-18} - 10^{-16} cm^3/s$, and hence for the
secondary electron density produced by the neutralino
annihilation. Therefore, we obtain that ICS emission expected from
these models is very large and exceeds both the EUVE and BeppoSAX
data as well as the EGRET and Fermi upper limits on Coma (the HESS
upper limit on Coma at very high energies cannot strongly
constrain the SED produced in the DM models that we consider
here). We conclude, therefore, that neutralino DM models with mass
of order of 500 GeV must be excluded by our analysis.

The shape of the SED does not change substantially when a cored DM
profile is considered. The only effect is that the cored DM profile
requires slightly larger values on $\sigma V$ to fit the Coma
radio halo data.

\begin{figure}[ht]
\begin{center}
 \epsfig{file=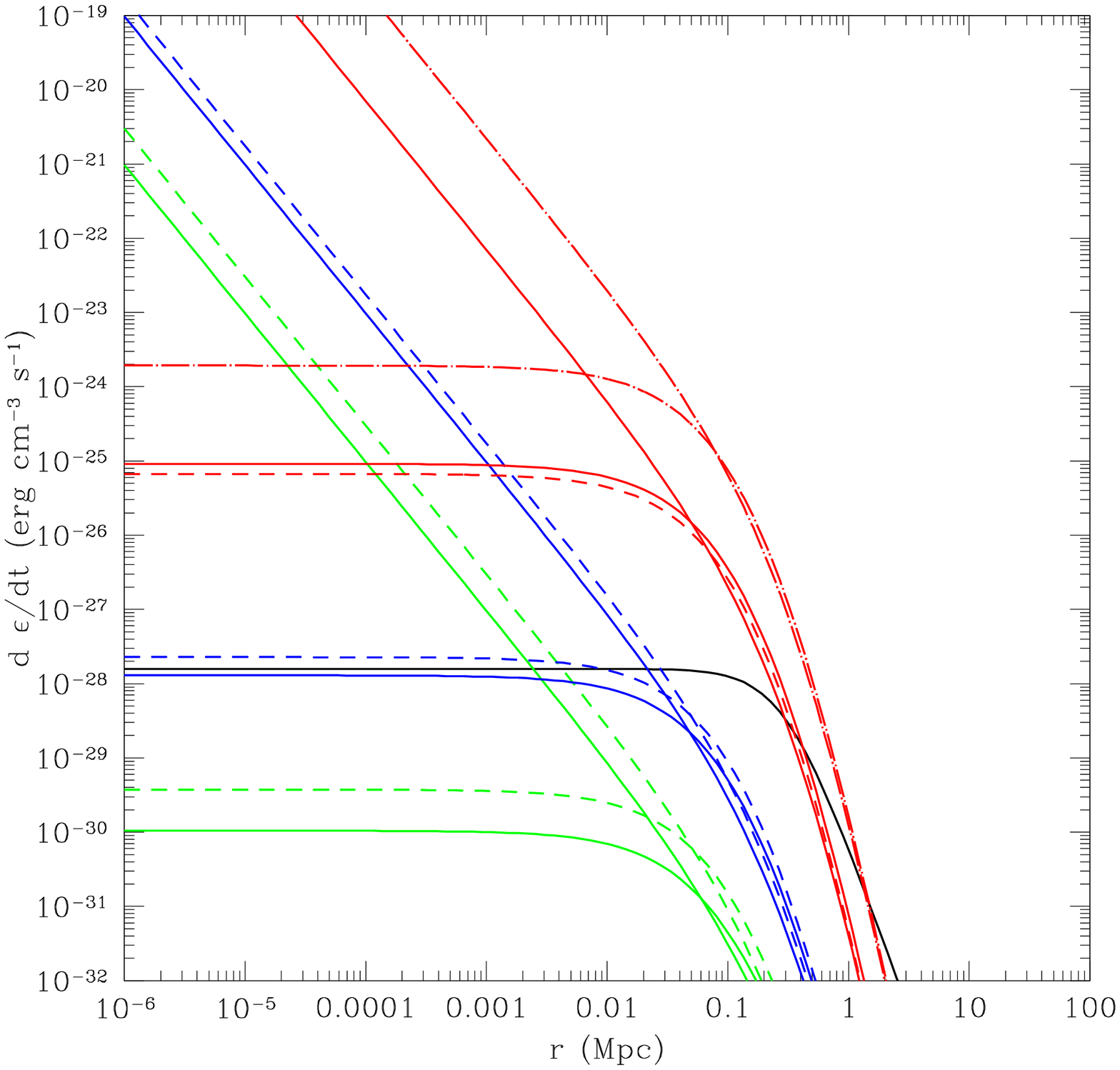,height=8.cm,width=9.cm,angle=0.0}
\end{center}
 \caption{\footnotesize{The heating rate induced by DM secondary particles in Coma
 as predicted by the three DM models with $M_{\chi}=9$ GeV, $M_{\chi}=60$ GeV and
 $M_{\chi}=500$ GeV for $b {\bar b}$ (solid), $\tau^{\pm}$ (dashed), W$^{\pm}$
 (dot-dashed) compositions.
 We consider an NFW DM profile (green, blue and red peaking curves for 9, 60 and 500 GeV respectively)
 and a cored DM profile (green , blue and red flattening curves  for 9, 60 and 500 GeV respectively).
 The values of $\sigma V$ and $B_0$ for the different models are
 the same used in Fig.\ref{Coma_rh_bestfit}.
 The solid black curve shows the bremsstrahlung
 cooling rate of the intra-cluster gas at a temperature of  $8.2$ keV.
 The results are obtained using $B(r)\propto n_{th}(r)$ and smooth DM profiles, without considering the effect of substructures.
 }}
 \label{Coma_heating}
\end{figure}

We also evaluate the intracluster gas heating induced by DM
annihilation (see Fig.\ref{Coma_heating}) and we find that an NFW
DM density profile produces an excess heating in the inner $\sim
3$ kpc and $\sim 30$ kpc regions of Coma for 9 and 60 GeV,
respectively. This effect provides, hence, a strong constraint to
the annihilation cross-section for these DM models. In order to
have the DM-induced heating rate lower than the bremsstrahlung
cooling rate at the cluster center the annihilation cross section
worked out here must be reduced by a factor $\sim 10^7 - 10^{12}$.

The heating rate for the case of a cored DM profile with a mass of
500 GeV is larger than the bremsstrahlung cooling rate of Coma for
every DM annihilation channel. Therefore, also this result
excludes the 500 GeV mass neutralino model.\\
DM models with a mass of 60 GeV and a cored profile show, instead,
a heating rate 
smaller than the bremsstrahlung cooling rate for the $b\bar b$
composition, and are therefore consistent with both
multi-frequency SED data and with heating rate constraints, while
the $\tau^\pm$ model is not consistent because it has a heating
rate greater than the cooling rate. For a neutralino mass of 9
GeV and cored DM profile, both $b\bar b$ and $\tau^\pm$ models are consistent with
heating rate constraints.

\section{Discussion and conclusions}

Any DM interpretation of astrophysical data has intrinsic
multi-frequency correlations and should, therefore, be tested
against a series of independent observational constraints.

The full SED induced by DM annihilation in Coma indicates that the
non-thermal e.m. signals induced by a 9 or 60 GeV neutralino can
be consistent with the shape of the radio-halo spectrum and
produce ICS, bremsstrahlung radiation and $\pi^0 \to \gamma
\gamma$ radiation consistent with the existing limits at
multi-frequency. However, these signals cannot, in particular, be
the nature of the soft and hard X-ray emission excesses detected
in the Coma cluster. An exception to this point is given by the
$\tau^\pm$ model with mass 60 GeV that is able to fit the hard
X-ray emission observed in Coma. However, in the framework of this
model, the EUV excess in Coma should then be interpreted as due to
thermal emission from warm baryons at sub-virial temperature.

Focusing on the radio frequency range,, we have shown that a DM
particle annihilating into $b \bar b$ is best fitting the spectral
shape of the radio halo of the Coma cluster, quite independently
of the neutralino mass.
However, the radio halo spectrum fit requires that
i) the value of the neutralino annihilation cross section $\sigma
V$ strongly increases with increasing neutralino mass, and that
ii) the best fit value of the central magnetic field in Coma
decreases with increasing neutralino mass.\\
In the 9 GeV case the value of the central magnetic field required
in the $b\bar b$ case ($B_0\sim20$ $\mu$G) is, however, too high
compared to the results of Faraday Rotation measures (Carilli \&
Taylor 2002, Bonafede et al. 2010), while in the $\tau^\pm$ case
an acceptable value of $B_0\sim2.2$ $\mu$G is required. The best
fit values of $B_0$ decrease then to still marginally acceptable
values $ \sim 0.1 - 0.5 $ $\mu$G for the 60 GeV case (still
consistent with the available observational limits, see Carilli \&
Taylor 2002), while they take extreme values as low as
$0.001-0.01$ $\mu$G for the highest neutralino mass (i.e. 500 GeV)
we consider in our study.
\begin{figure}[ht]
\begin{center}
{
 \epsfig{file=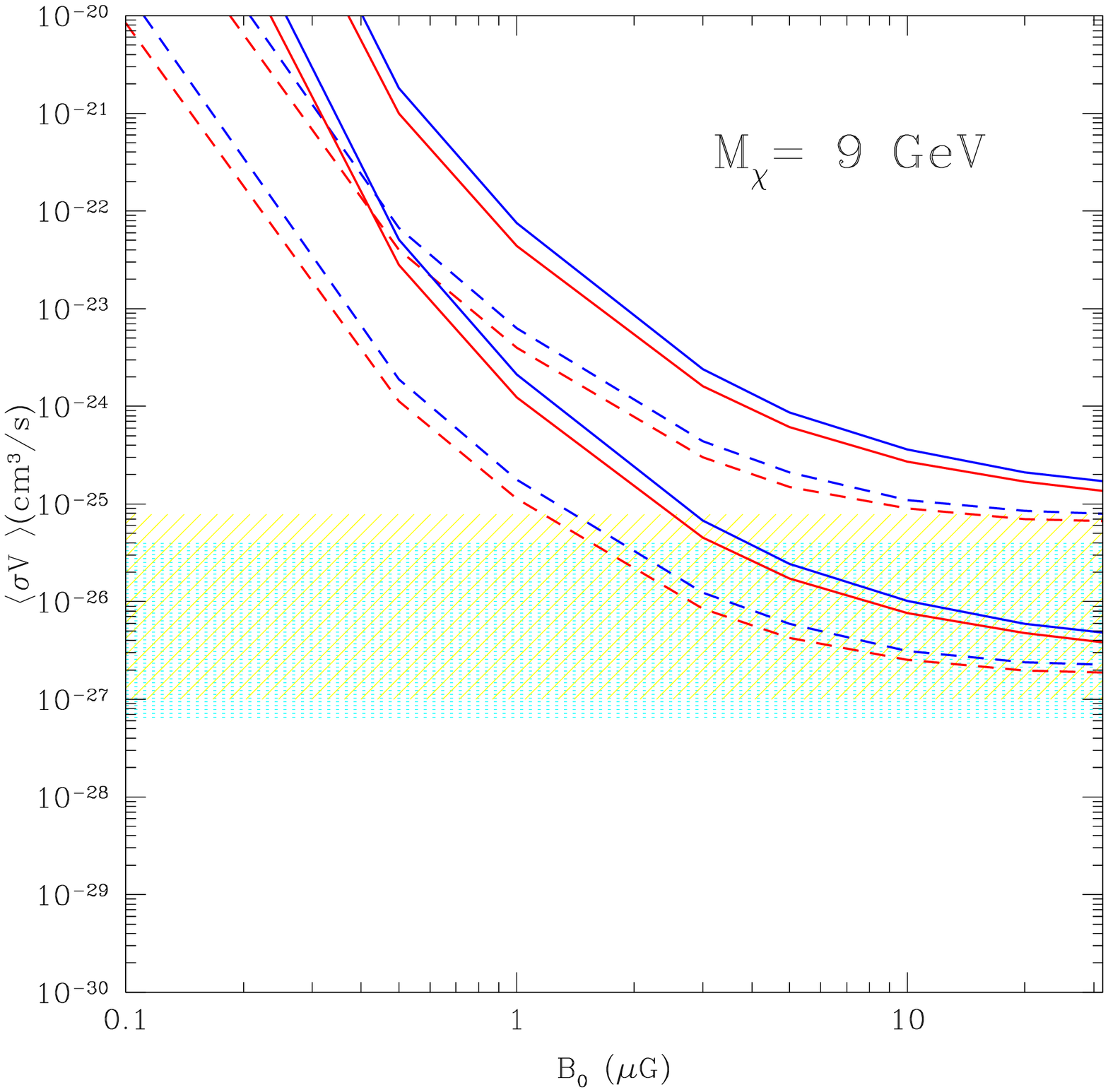,height=6.cm,width=8.cm,angle=0.0}
 \epsfig{file=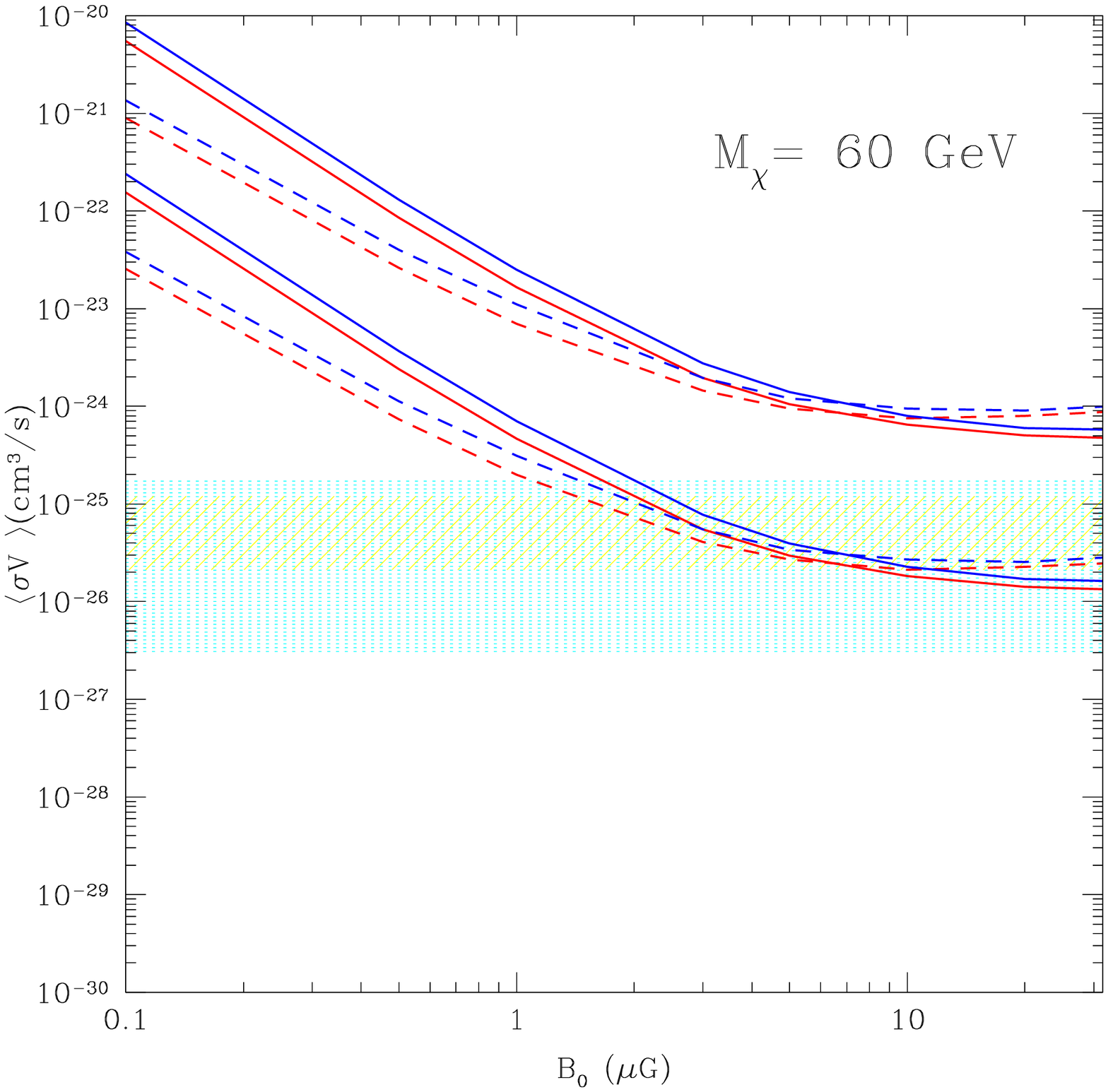,height=6.cm,width=8.cm,angle=0.0}
 \epsfig{file=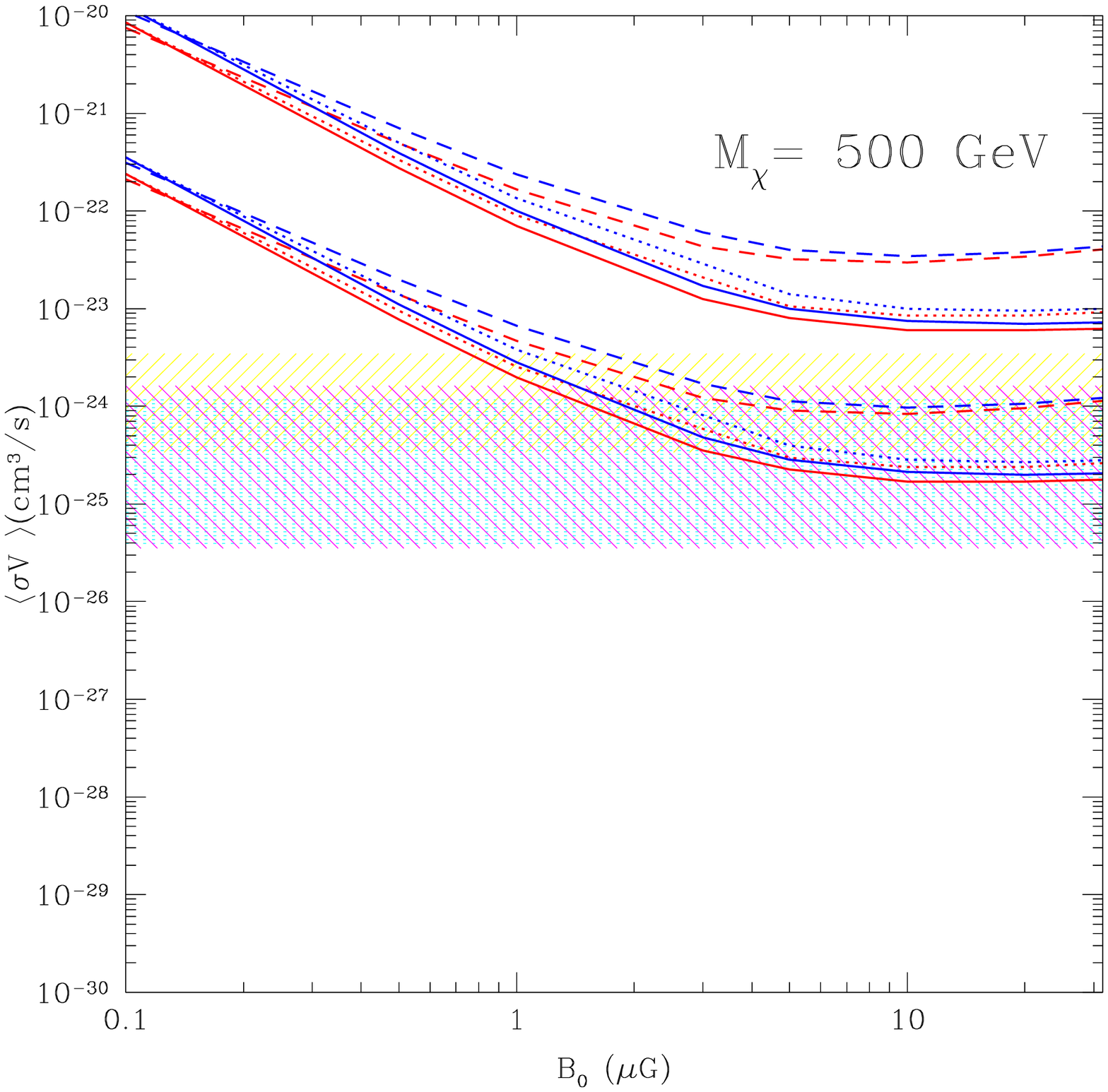,height=6.cm,width=8.cm,angle=0.0}
}
\end{center}
 \caption{\footnotesize{We show the constraints set by the radio halo spectrum
 of Coma on the annihilation cross section for the DM models considered
in this paper. Upper, middle and lower panel refer to the 9, 60
and 500 GeV neutralino mass, respectively. We consider the case of
an NFW DM profile with
 $b {\bar b}$ (solid red curve) and $\tau^{\pm}$ (dashed red
 curve), and the case for a cored DM profile with $b {\bar b}$ (solid blue curve)
 and $\tau^{\pm}$ (dashed blue curve). The lower panel contains also the
 cases of W$^{\pm}$ composition with NFW (dotted red curve) and cored DM profile
 (dotted blue curve).
 The upper curve bundles refer to a smooth DM profile while the
 lower curve bundles refer to the case in which substructures are
 considered as in Fig.\ref{Coma_boost}.
 The shaded bands in this figure represent the upper limit to the annihilation
 cross-section for the DM models derived combining the limits coming from a
 multi-messenger analysis (see text for details).
 The results are obtained using $B(r)\propto n_{th}(r)$.
 }}
 \label{sigmav_b0}
\end{figure}
\begin{table*}
\centering
\begin{tabular}{ll|c|c|c}
\hline \hline
 \multicolumn{5}{c}{upper limit on the annihilation cross-section ($10^{-26}\textrm{ cm}^3\textrm{s}^{-1}$)} \\
\hline
 \multicolumn{2}{c|}{DM benchmark} & VL2  & Aq & iso \\
\hline $M_{\chi}=9\textrm{ GeV}$ & $\chi\chi\to \tau^{+}\tau^{-}$
& $(\bf{0.1},\bf{0.1},\bf{0.1})$ (rad,rad,rad)
&$($$\bf7.8$$,$$\bf7.8$$,$$\bf7.8$$)$ (opt,opt,opt)
&$($$\bf7.8$$,$$\bf7.8$$,$$\bf7.8$$)$ (opt,opt,opt) \\
$M_{\chi}=9\textrm{ GeV}$ & $\chi\chi\to b\bar{b}$ &
$(\bf{0.06},\bf{0.06},\bf{0.06})$ (rad,rad,rad)
&$($$\bf4.3$$,0.7,0.2)$ (opt,anp,anp) &$($$\bf4.3$$,2.6,0.8)$
(opt,anp,anp) \\ $M_{\chi}=60\textrm{ GeV}$ & $\chi\chi\to
\tau^{+}\tau^{-}$ & $(2.2,2.2,2.2)$ (rad,rad,rad)
&$(5.3,3.6,\bf{2.1})$ (pos,pos,pos) &$($$\bf12$$,8.5,8.0)$
(pos,pos,pos) \\ $M_{\chi}=60\textrm{ GeV}$ & $\chi\chi\to
b\bar{b}$ &         $(0.6,0.6,0.6)$ (rad,rad,rad)
&$(8.5,0.8,\bf{0.3})$ (anp,anp,anp) &$($$\bf19$$,3.8,1.7)$
(anp,anp,anp) \\ $M_{\chi}=500\textrm{ GeV}$ & $\chi\chi\to
\tau^{+}\tau^{-}$ &$(106,101,73)$ (rad,pos,pos) &
$(125,64,\bf{34})$ (GCg,pos,pos) &  $($$\bf349$$,184,170)$
(pos,pos,pos) \\ $M_{\chi}=500\textrm{ GeV}$ & $\chi\chi\to
W^{+}W^{-}$ &      $(20,20,7.0)$ (rad,rad,anp) &
$(74,8.6,\bf{3.5})$ (pos,anp,anp) & $($$\bf162$$,46,20)$
(pos,anp,anp) \\ $M_{\chi}=500\textrm{ GeV}$ & $\chi\chi\to
b\bar{b}$ &        $(12,12,7.4)$ (rad,rad,anp) &
$(54,9.2,\bf{3.7})$ (pos,anp,anp) & $($$\bf118$$,49,21)$
(pos,anp,anp) \\ \hline \hline
\end{tabular}

\caption{\fontsize{9}{9}\selectfont The upper limits on the
annihilation cross-sections in units of $10^{-26}\textrm{
cm}^3\textrm{s}^{-1}$ from several measurements for the seven
benchmark DM models. Each column shows the limits derived using
different galactic DM distributions: Via Lactea II (VL2), Aquarius
(Aq) and a cored isothermal profile (iso). Each cell features
three values corresponding to the minimal, mean and maximal
propagation setups respectively. Also shown is the channel that
produces the quoted upper limit $-$ ``rad'', ``pos'', ``anp'',
``GCg'' and ``opt'' stand for radio, positrons, antiprotons,
Galactic Centre $\gamma$-rays and CMB optical depth, respectively.
The extreme values for each benchmark are indicated in bold and
delimit the shaded regions of Fig.\ref{sigmav_b0}.}
 \label{tab2}

\end{table*}
We have also studied the variation of $\sigma V$ for increasing
values of $B_0$ that could still provide reasonable fit to the
Coma radio halo spectrum. Fig.\ref{sigmav_b0} shows that the
curves in the $\sigma V - B_0$ plane that are locus of the best
fit to the Coma radio halo flatten to approximately the same value
of $\sigma V$ for high values of $B_0$, i.e. when the energy
losses are dominated by synchrotron losses.

The shaded bands in Fig.~\ref{sigmav_b0} encompass the range of
upper limits on the annihilation cross-section derived from
multi-messenger constraints. Such constraints, summarised in Table
\ref{tab2}, include positrons, antiprotons, radio emission and
$\gamma$-rays from the Galactic Centre
and the optical depth of CMB photons. The uncertainty related to
the propagation of antimatter in our Galaxy is taken into account
by using minimal, mean and maximal propagation setups as in Donato
et al. (2004). Galactic DM distribution is modelled according to
the smooth and clumpy distributions found in the latest
high-resolution $N$-body simulations, namely Via Lactea II
(Diemand et al. 2008) and Aquarius (Springel et al. 2008). As a
conservative case, also a cored isothermal profile with no
substructure is considered. For further details we refer to Pato
et al. (2009), Pieri et al. (2009) and Catena et al. (2009).
We stress that these multi-messenger bounds are quite robust and
not easily avoidable, and therefore they set further constraints
to the seven benchmark DM models studied in the present work (see
Fig.~\ref{sigmav_b0}). Other upper limits on the annihilation
cross-section may be derived e.g.~from the observation of dwarf
galaxies with the Fermi satellite (Farnier 2009; see Fig.3 of Abdo
et al. 2010), which exclude values greater than $\sim
10^{-25}\textrm{ cm}^3\textrm{s}^{-1}$ for a neutralino mass of 60
GeV. However, taking into account the uncertainty in the density
profile as well as the dependence of the Fermi-LAT sensitivity
from the assumed source spectrum (which is rather steep for the
case of the neutralino DM annihilation), such a limit could be
released toward slightly higher values.\\
The flattening of the curves in the $\sigma V - B_0$ plane for
high values of $B_0$ means that it is not possible to have values
of $\sigma V$ that match the allowed region of annihilation cross
sections suggested by multi-messenger studies of DM annihilation
simply by increasing the value of the central magnetic field in
galaxy clusters.

The lower set of curves in Fig.\ref{sigmav_b0} shows the
annihilation cross sections required to fit the radio halo
spectrum if DM substructures in Coma are considered (according to
our discussion in Sect. 2). The effect of substructures is to
increase the surface brightness intensity of the radiation emitted
by the electrons, especially at large radii. As shown in
Fig.\ref{sigmav_b0}, the inconsistency between fitting the radio
halo and the multi-messenger analysis is significantly improved,
particularly for $B_0 \simgt 1 \mu G$. However, none of the DM
models under study is safely below the annihilation cross section
upper limits obtained from the multi-messenger constraint analysis
whose results are summarized in Tab.\ref{tab2}.

We have also considered the predictions of the various DM models
by using a B-field of Coma that has the intensity and radial
profile suggested by recent Faraday Rotation measures (Bonafede et
al. 2010, see Fig. \ref{Coma_rh}). In this case, the DM model with
60 GeV and $b\bar b$ composition and the model with 9 GeV and
$\tau^\pm$ composition would produce a radio halo spectrum with a
shape that is still consistent with the observed one.
The value of the annihilation cross section required in the 60
GeV, $b\bar b$ case is $\sigma V=1.0\times10^{-24}$ cm$^3$
s$^{-1}$, while in the 9 GeV, $\tau^\pm$ case the cross section
required is $\sigma V=1.4\times10^{-25}$ cm$^3$ s$^{-1}$.\\
However, in order to have a DM-induced heating rate lower than the
bremsstrahlung cooling rate of the intracluster gas at the cluster
center, the annihilation cross sections worked out here must be
reduced by a factor $\sim 10^7 - 10^{12}$ for $M_{\chi}$ from 9 to
500 GeV, if we assume an NFW density profile.

This problem is, however, not present for a neutralino mass of 9
GeV (with both $b\bar b$ and $\tau^\pm$ compositions) and 60 GeV
and composition $b\bar b$ if DM density has a cored profile (see
Fig.\ref{Coma_heating}), while it still holds for the 60 GeV mass
with composition $\tau^\pm$ and for all neutralino compositions in
the 500 GeV mass model.

The high-mass DM model with $M_{\chi}=500$ GeV, whose SED is
normalized in order to fit the Coma radio-halo data with a very
low magnetic field value ($B_0 \approx 0.01$ $\mu$G for the $b\bar
b$ composition), is inconsistent with the multi-frequency SED
because it exceeds by large factors the EUV, soft X-ray, hard
X-ray and gamma-ray (in the energy range of Fermi) upper limits
(see Fig.\ref{Coma_sed}).
For such high-mass DM model, also the $\tau^{\pm}$ and $W^{\pm}$
cases barely fail to fit the Coma radio-halo spectrum at low
frequency below $\sim 400$ MHz (see Fig.\ref{Coma_rh_bestfit}).

We note finally that the high-energy HESS gamma-ray limit is not
relevant in any of the DM models considered in this paper, because
in all the cases the threshold energy of the HESS upper limits is
larger than the neutralino masses here considered.

As a final remark, we stress that the DM candidates with $M_{\chi}=9$
GeV and $M_{\chi}=60$ GeV -- the ones that best fit Coma astrophysical
data --  are too light to accommodate the PAMELA positron fraction
(Adriani et al 2009) or the electron plus positron flux seen by Fermi-LAT
(Abdo et al. 2009) and HESS (Aharonian et al. 2008, 2009b).

In conclusion, the possibility of interpreting the origin of
non-thermal phenomena in galaxy clusters with DM annihilation
scenarios would require a low or intermediate neutralino mass and
a cored DM density profile. If we then consider the multimessenger
constraints to the neutralino annihilation cross-section, it turns
out that such scenarios would also be excluded unless we introduce
a substantial boost factor due to the presence of DM
substructures.

\begin{acknowledgements}
MP acknowledges a grant from Funda\c{c}\~{a}o para a Ci\^encia e
Tecnologia (Minist\'erio da Ci\^encia, Tecnologia e Ensino
Superior).
LP thanks the University of Trento for hospitality.
\end{acknowledgements}

\end{document}